\documentclass[aps,pra,twocolumn,amsmath,amssymb,superscriptaddress,reprint,longbibliography]{revtex4-1}

\usepackage[utf8]{inputenc}
\usepackage{graphicx}
\usepackage[colorlinks=true,citecolor=blue]{hyperref}
\usepackage{units}
\usepackage{amsmath}

\DeclareMathOperator{\re}{\text{Re}}
\DeclareMathOperator{\im}{\text{Im}}

\AtBeginDocument{%
    \newwrite\bibnotes
    \def\bibnotesext{Notes.bib}
    \immediate\openout\bibnotes=\jobname\bibnotesext
    \immediate\write\bibnotes{@CONTROL{REVTEX41Control}}
    \immediate\write\bibnotes{@CONTROL{%
    apsrev41Control,author="08",editor="1",pages="1",title="0",year="1"}}
     \if@filesw
     \immediate\write\@auxout{\string\citation{apsrev41Control}}%
    \fi
}%

\begin{document}

\title{Nonlinear chiral refrigerators} 

\author{David S\'anchez}
\affiliation{Institute for Cross-Disciplinary Physics and Complex Systems IFISC (UIB-CSIC), E-07122 Palma de Mallorca, Spain}
\author{Rafael S\'anchez}
\affiliation{Departamento  de  F\'{\i}sica  Te\'orica  de  la  Materia  Condensada
and Condensed  Matter  Physics  Center  (IFIMAC),
Universidad  Aut\'onoma  de  Madrid,  28049  Madrid,  Spain}
\author{Rosa L\'opez}
\affiliation{Institute for Cross-Disciplinary Physics and Complex Systems IFISC (UIB-CSIC), E-07122 Palma de Mallorca, Spain}
\author{Bj\"{o}rn Sothmann}
\affiliation{Theoretische Physik, Universit\"at Duisburg-Essen and CENIDE, D-47048 Duisburg, Germany}


\begin{abstract}
We investigate a mesoscopic refrigerator based on chiral quantum Hall edge channels.
We discuss a three-terminal cooling device in which charge transport occurs between a pair of voltage-biased terminals only.
The third terminal, which is to be cooled, is set as a voltage probe with vanishing particle flux.
This largely prevents the generation of direct Joule heating which ensures a high coefficient of performance.
Cooling operation is based on energy-dependent quantum transmissions.
The latter are implemented with the aid of  two tunable scattering resonances (quantum dots).
To find the optimal performance point and the largest temperature difference created with our refrigerator, it is crucial to address the nonlinear regime of transport, accounting for electron-electron interaction effects.
Our numerical simulations show that the maximal cooling power can be tuned with the quantum dot couplings and energy levels.
Further, we provide analytical expressions within a weakly nonlinear scattering-matrix formalism which allow us to discuss the conditions for optimal cooling in terms of generalized thermopowers.
Our results are important for the assessment of chiral conductors as promising candidates for efficient quantum refrigerators with low dissipation.
\end{abstract}


\maketitle


 \section{Introduction}\label{sec:intro}
A great obstacle for the operation of nanoscale integrated circuits is heat dissipation, which can dramatically alter their performance. Thus, it is highly desirable to have nanoelectronic devices where heat can be removed or converted into useful electric work. To this purpose, thermoelectric phenomena can be exploited by taking into account {\em both} particle and energy fluxes. As a matter of fact, the field of quantum thermoelectrics has experienced a rapid development in the last two decades using low-dimensional nanodevices~\cite{benenti_fundamental_2017}, such as quantum
dots~\cite{staring_coulomb-blockade_1993,dzurak_observation_1993,dzurak_thermoelectric_1997,scheibner_thermopower_2005,scheibner_sequential_2007,nakpathomkun_thermoelectric_2010,thierschmann_diffusion_2013,svensson_nonlinear_2013,josefsson_quantum-dot_2018}. A thermoelectric device can use electrical power to extract heat from a reservoir, hence operating as a refrigerator (Peltier effect) or, by reciprocity, transform a temperature difference into an output power, thus acting as a heat engine (Seebeck effect). When nanoscale heat engines are connected to multiple terminals, they behave as powerful and efficient energy harvesters~\cite{sothmann_thermoelectric_2015}. This is because the directions of charge and heat currents become decoupled~\cite{sanchez_optimal_2011}. The effect have been experimentally demonstrated with three-terminal quantum-dot setups~\cite{roche_harvesting_2015,hartmann_voltage_2015,thierschmann_three-terminal_2015,jaliel_experimental_2019}, paving the way for a new generation of multiterminal, highly efficient nanoscale heat engines. In the quantum Hall regime due to strong magnetic fields, the conducting channels are chiral and precisely the nature of edge states gives rise to powerful energy harvesters~\cite{sothmann_quantum_2014,sanchez_chiral_2015} and ideal thermal diodes~\cite{sanchez_heat_2015}. Quite generally, broken time-reversal symmetry facilitates the appearance of large efficiencies at finite output power~\cite{brandner_strong_2013,brandner_multi-terminal_2013,brandner_bound_2015} due to symmetry breaking of the off-diagonal elements of the Onsager matrix~\cite{sanchez_thermoelectric_2011,saito_thermopower_2011}. Surprisingly, the operation of such devices in their dual role as refrigerators has thus far been largely unexplored. Recent experiments, however, emphasize the thermoelectric properties of chiral edge states~\cite{granger_observation_2009,nam_thermoelectric_2013}, and permit to trace the relaxation of energy along the propagation channels~\cite{altimiras_non-equilibrium_2010,altimiras_tuning_2010,ota_spectroscopic_2019,rodriguez_strong_2019}.

\begin{figure}[b]
    \includegraphics[width=.85\linewidth]{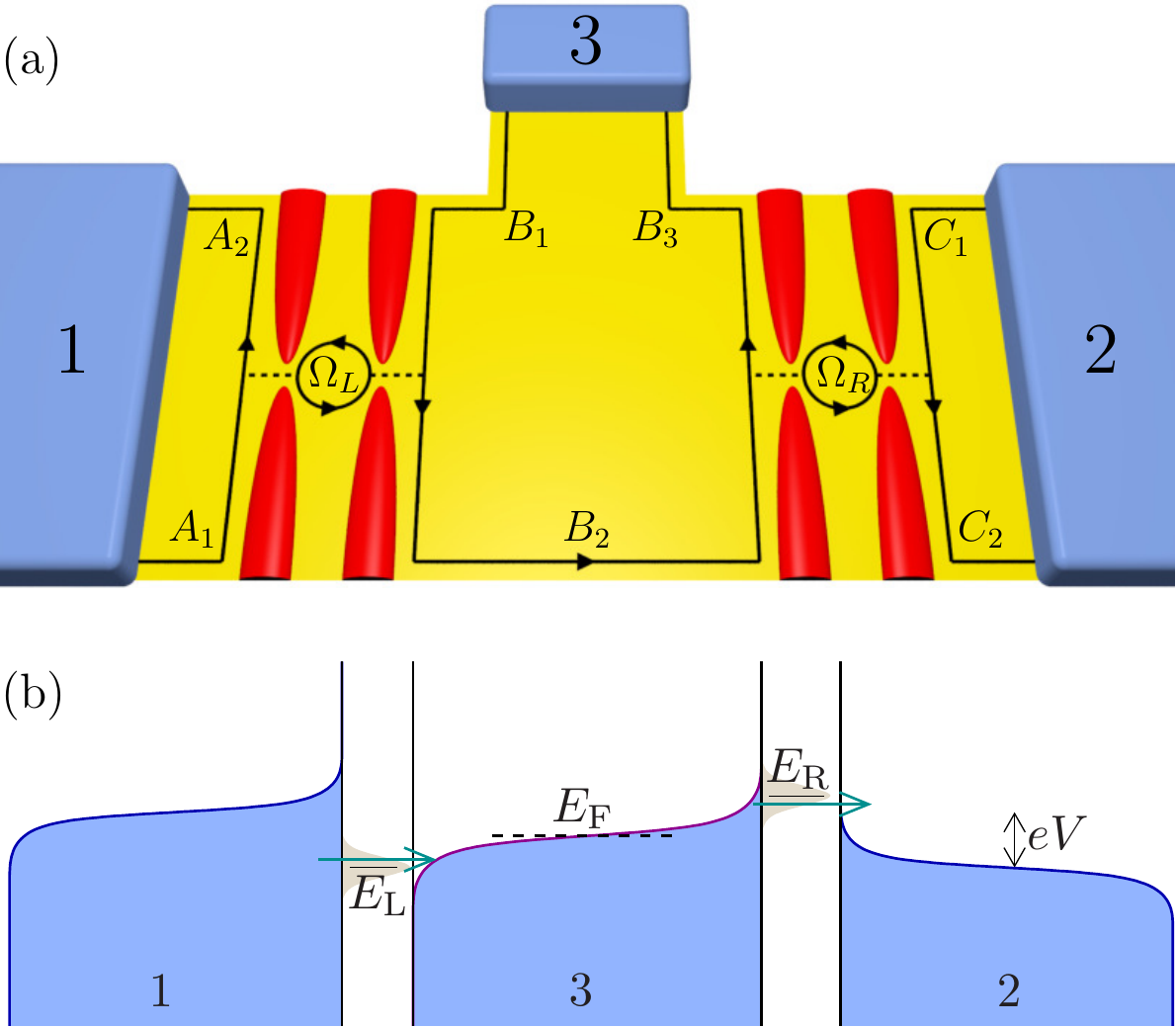}
	\caption{\label{fig:nonlinear} (a) Schematic of the model system used for the analysis of the nonlinear chiral refrigerator proposal. The edge channels are split into different regions $A_i$, $B_i$ and $C_i$ which are assumed to have a constant electrostatic potential.
	(b) Energy diagram for the tunneling electrons across the two quantum dots ($L$ and $R$) in the operational mode when heat is extracted from terminal 3.}
\end{figure}

While there have been a few proposals of two-terminal refrigerators based on resonant tunneling through quantum dots~\cite{edwards_quantum-dot_1993,edwards_cryogenic_1995,prance_electronic_2009,gasparinetti_probing_2011}, superconductor-normal metal junctions~\cite{nahum_electronic_1994,leivo_efficient_1996,giazotto_opportunities_2006,muhonen_micrometre-scale_2012} and metallic islands and quantum dots in the Coulomb-blockade regime~\cite{timofeev_electronic_2009,arrachea_heat_2007,rey_nonadiabatic_2007,cleuren_cooling_2012,bruggemann_cooling_2014,pekola_refrigerator_2014}, examples of multiterminal refrigerators comprise minimal models of refrigerators consisting of qubits coupled to bosonic baths~\cite{linden_how_2010,brunner_virtual_2012,levy_quantum_2012,brunner_entanglement_2014,correa_quantum-enhanced_2014,correa_multistage_2014}, arrangements of four quantum dots~\cite{venturelli_minimal_2013} and mixed junctions~\cite{entin-wohlman_enhanced_2015,sanchez_correlation-induced_2017,sanchez_cooling_2018,hussein_nonlocal_2019}.
In this paper, we investigate a three-terminal chiral refrigerator based on quantum Hall edge states.

A sketch of the proposed device is shown in
Fig.~\ref{fig:nonlinear}(a). Our setup is interesting for a number of reasons. The system is based on transport along a single quantum channel. Thus, it potentially allows for much larger cooling powers than refrigerators based on tunnel junctions. Importantly, the chiral refrigerator is able to cool a terminal
which is several micrometers away due to the fact that edge states can travel long distances
with very little dissipation. Additionally, the third terminal which is to be cooled carries no charge current. As a consequence, there is no direct Joule heating and this allows us to achieve highly efficient coolings without heating the electronic environment,
which is an appealing feature from the experimental perspective. As there is no direct Joule heating, it is crucial to include nonlinear effects in the theoretical analysis of the refrigerator in order to properly estimate the maximal cooling power and the minimal reachable temperature~\cite{whitney_nonlinear_2013}. We show that both power and efficiency of our  quantum dot-based refrigerator can be tuned with either the width of the dot levels or their positions. Both parameters can be experimentally manipulated with electric fields applied to nearby gate terminals. All our findings are obtained employing the generalized scattering theory formalism for treating thermoelectric transport in which we include interactions to properly describe the nonlinear regime \cite{,sanchez_nonlinear_2016}.

\section{The system}

We consider a three-terminal conductor in the quantum Hall regime with filling factor $1$. Therefore, only the lowest Landau level is occupied and there exists a single chiral edge state running along the boundary of the sample. This is represented in Fig.~\ref{fig:nonlinear}(a) with solid lines and arrows indicating the propagation direction. The basic operation mechanism consists of driving a charge current between terminals $1$ and $2$ by applying a bias voltage $V_1-V_2$.
The device also needs some type of energy-dependent scattering. This is accomplished by inserting quantum scatterers near terminals $1$ and $2$. For instance, one can define negatively biased top-gate fingers [red areas in Fig.~\ref{fig:nonlinear}(a)] on top of the electron gas (yellow area). For concreteness, in the following we take these scatterers to be quantum dots $r=L,R$, each having a single resonant level with energy $E_r$ and tunneling-induced level width $\Gamma_r$. 
Electrons injected from $1$ tunnel through the left resonance and occupy states in $3$ below $E_F$ [Fig.~\ref{fig:nonlinear}(b)].
At the same time, thermally excited electrons in $3$ tunnel through the right resonance and occupy states in $2$ above $E_F$. Since the voltage $V_3$ adjusts itself such that charge current into terminal 3 vanishes, the overall effect
is a heat extraction from terminal 3 since states below (above) the Fermi energy become progressively populated (depopulated)
while the particle current is kept to zero.
In an experiment, the latter condition is easily achieved by operating terminal 3 as a floating contact (voltage probe).
We emphasize that the working principle of our thermoelectric refrigerator takes advantage on chirality---a preferred propagation direction of electronic motion.  
Chirality can be exploited to generate spin polarizations~\cite{gohler_spin_2011}, 
departures of the Onsager reciprocity~\cite{sanchez_chirality_2005}, spin caloritronic effects~\cite{lopez_fluctuation_2012,hwang_nonlinear_2014},
heat rectification~\cite{vannucci_interference-induced_2015}, and spin refrigeration~\cite{mani_helical_2018,roura-bas_helical_2018}.
Our proposal is unique in the sense that it combines all-electrical chirality and nonlinearity to achieve maximum cooling power. 

\section{Theory}

In order to discuss the nonlinear response of the refrigerator, we employ the nonlinear scattering matrix theory for charge and heat transport~\cite{sanchez_scattering_2013,meair_scattering_2013,lopez_nonlinear_2013}. At very low temperature and for conductors free of disorder, transport is elastic and the scattering matrix is a function of both the energy $E$ and the electron-electron potential landscape $U(\vec{r})$ of the conductor. The evaluation of the electrostatic profile requires to solve the Poisson equation, which is computationally costly. Instead, we here follow a simplified approach and split the potential into nine different regions $X=\{A_1,A_2,\Omega_L,B_1,B_2,B_3,\Omega_R,C_1,C_2\}$ as indicated in Fig.~\ref{fig:nonlinear}(a). In each
zone the potential is taken to be constant, $U_X$. The charge in region $X$ can be expressed as
\begin{equation}\label{qX}
	q_{X}=e\int dE \sum_j\nu_{X,j}(E)f_j(E) \,,
\end{equation}
where $\nu_{X,j}^p$ is the particle injectivity of lead $j=1,2,3$ in region $X$.
The injectivity is the density of states associated to those carriers originated from lead $j$.
In App.~\ref{sec:discretization} we give the full expressions for all these functions in the considered regions.
In Eq.~\eqref{qX}, $f_j(E)=1/\{\exp[(E-\mu_j)/ k_B T_j]+1\}$ is the Fermi function
with $\mu_j = E_F + eV_j$ the electrochemical potential of terminal $j$ and $T_j=T+\Delta T_j$ its temperature ($T$ is the base temperature
and $\Delta T_j$ the thermal shift).
Then, the net injected charge is $\delta q_{X}=q_{X}-q_{X}^{\rm eq}$, where $q_{X}^{\rm eq}$ is calculated from Eq.~(\ref{qX}) considering that all external dc potentials $V_{j}$ are equal and no temperature bias is applied ($\Delta T_j=0$).

Interaction between charges in different regions is described by a geometric capacitance matrix $\mathcal{C}_{X,X'}$.
According to the discretized version of the Poisson equation we have
\begin{equation}
\label{qXC}
	q_X=\sum_{X'} C_{X,X'} U_{X'}\,.
\end{equation}
We consider the strongly interacting limit, which is the experimentally relevant case. Therefore, we set $C_{X,X'}=0$. Thus,
Eqs.~(\ref{qX}) and~(\ref{qXC}) determine the internal potential at both dot sites.
We find for the injected charge in the two quantum dots:
\begin{align}
\delta q_{\Omega L}&=\frac{e\Gamma_L}{4\pi}\Big[\mathcal{I}_1^{L}(V_1,T_1)+\mathcal{I}_1^{L}(V_3,T_3)-2\mathcal{I}_{1,{\rm eq}}^{L}\Big]\,,
\\
\nonumber
\delta q_{\Omega R}&=\frac{e\Gamma_R}{4\pi}\Big\{\frac{\Gamma_L^2}{4}[\mathcal{I}_2(V_1,T_1) - \mathcal{I}_2(V_3,T_3)]
\\
&+\mathcal{I}_1^{R}(V_3,T_3)+\mathcal{I}_1^{R}( V_2,T_2)-2\mathcal{I}_{1,{\rm eq}}^{R}\Big\}\,,
\end{align}
where the integrals $\mathcal{I}_1^{r}$ and $\mathcal{I}_2$ are defined and evaluated in App.~\ref{sec:integrals}.
Note that for the equilibrium values we assume $U_{\Omega L}=U_{\Omega R}=0$, which can always be achieved by redefining the level positions $E_L$ and $E_R$.

Once the potential of each region is calculated, the expressions for the heat and charge currents in lead  $i$ are computed via the scattering matrix approach:
\begin{align}
	I^e_i&=\frac{e}{h}\sum_j\int dE \left[\delta_{ij}-\mathcal T_{i\leftarrow j}(E,U_X)\right]f_j(E),\\
	I^E_i&=\frac{1}{h}\sum_j\int dE E\left[\delta_{ij}-\mathcal T_{i\leftarrow j}(E,U_X)\right]f_j(E),\\
	I^h_i&=\frac{1}{h}\sum_j\int dE (E-\mu_i)\left[\delta_{ij}-\mathcal T_{i\leftarrow j}(E,U_X)\right]f_j(E),
\end{align}
where $\mathcal T_{i\leftarrow j}(E,U_X)$ denotes the energy-dependent transmission probability from terminal $j$ to terminal $i$. 
The currents are then expressed in terms of the integrals provided in App.~\ref{sec:integrals}. The sign convention for the fluxes is such that
heat (and particle) currents are positive when the flow is directed toward the sample. Therefore, positive values of $I^h_i$ imply {\em cooling}
of reservoir $i$.

\section{\label{sec:results}Results and discussion}

\begin{figure}
	\centering
	\includegraphics[width=.9\linewidth]{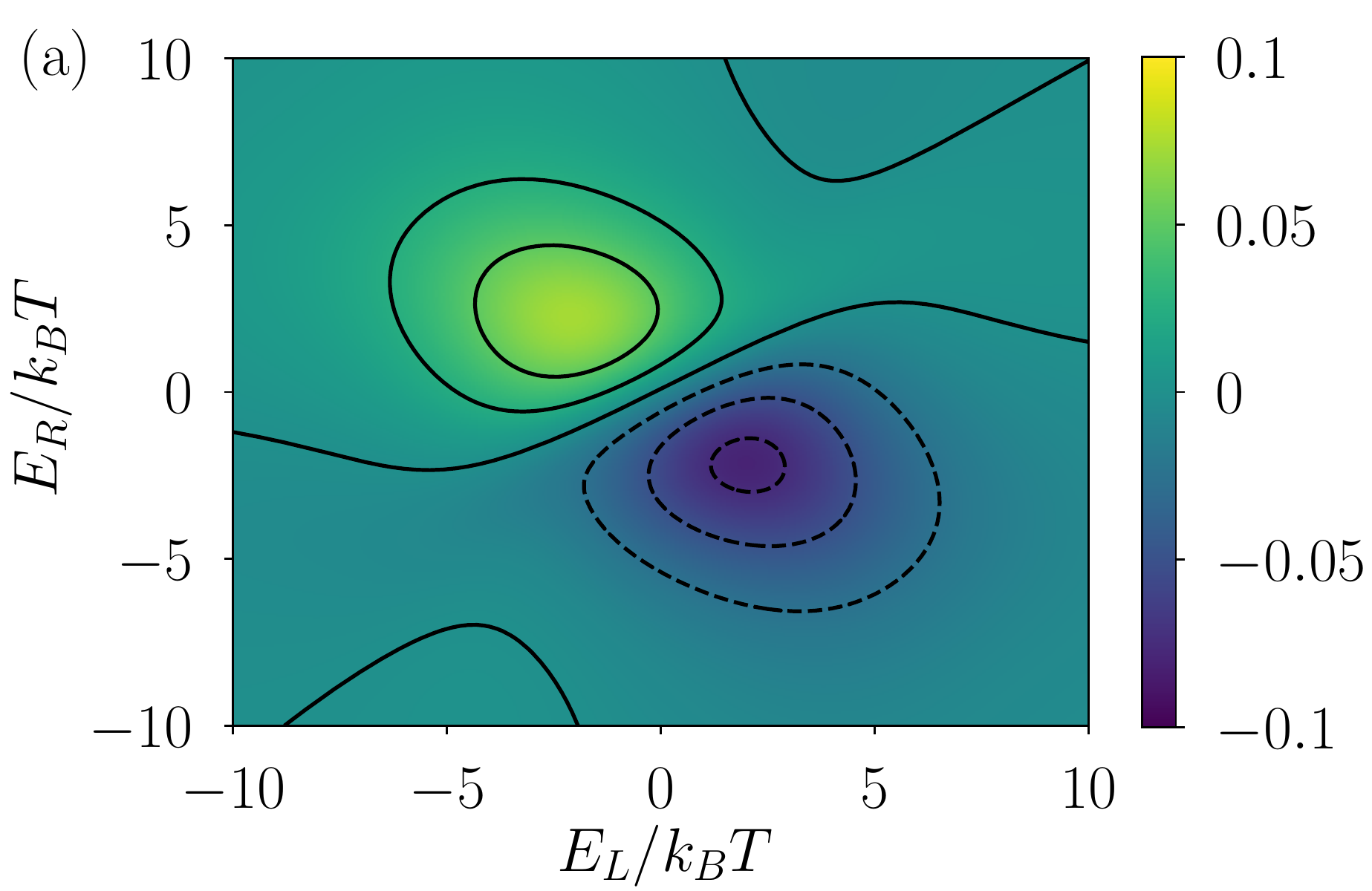}
	\includegraphics[width=.9\linewidth]{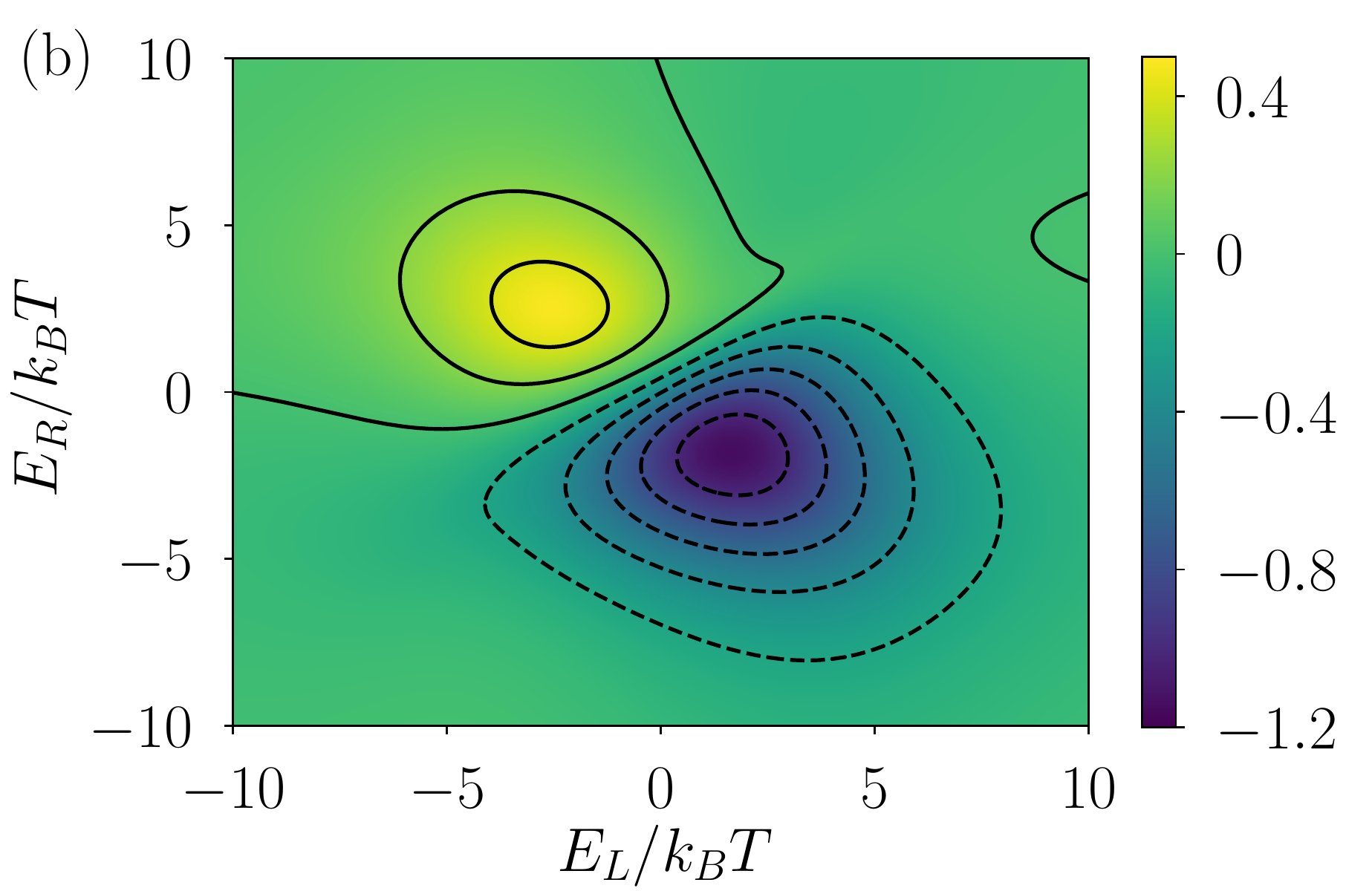}
	\includegraphics[width=.9\linewidth]{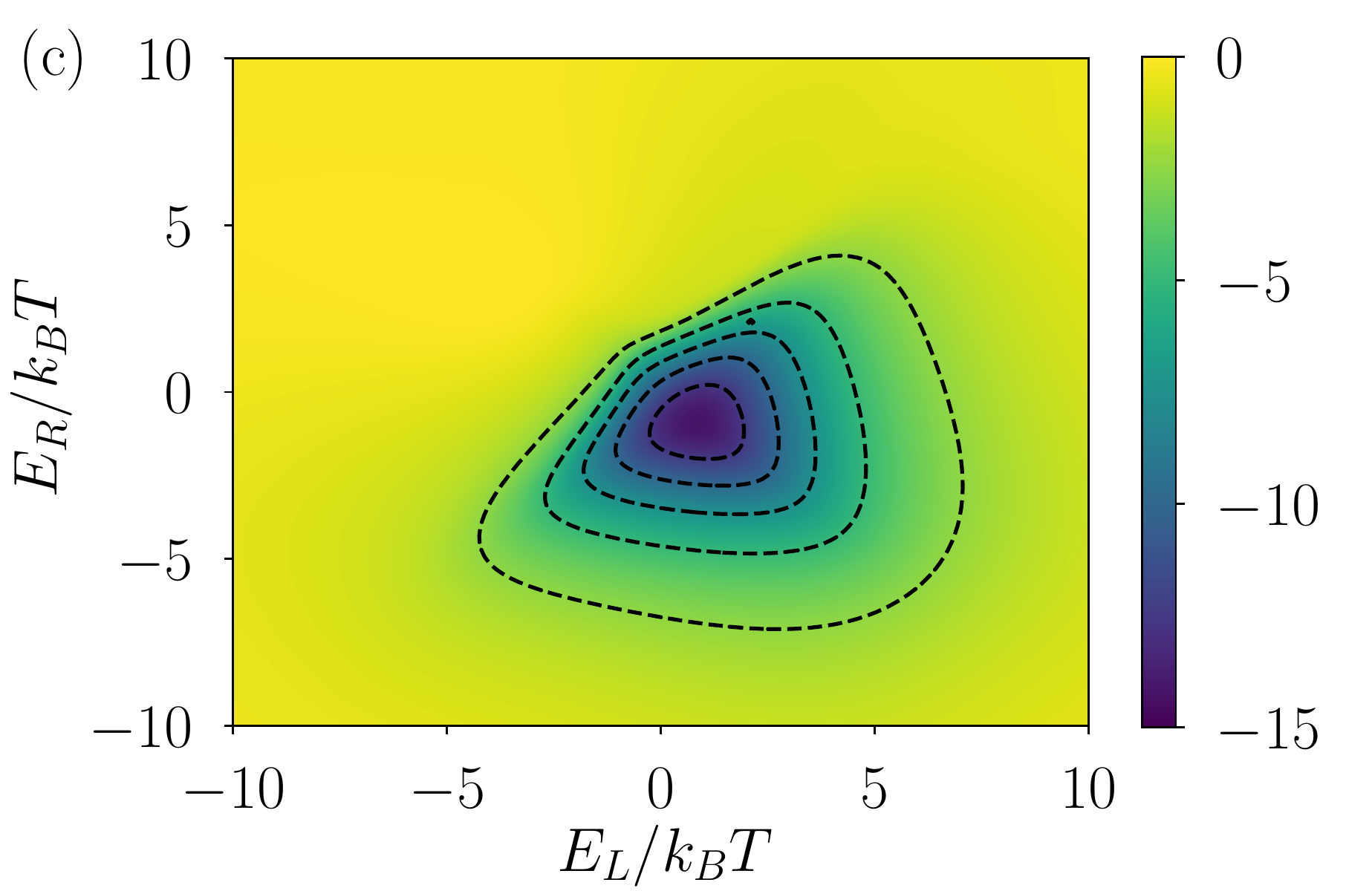}
	\caption{\label{fig:NLHeatCurrent}Heat current out of the third terminal $I^h_3$ [in units of $(k_B T)^2/h$]
	as a function of the two level positions for (a) $eV_1=-eV_2=0.1k_BT$, (b) $eV_1=-eV_2=k_BT$, and (c) $eV_1=-eV_2=5k_BT$. Parameters are $\Gamma_L=\Gamma_R=\Gamma = 2.6k_BT$, $\Delta T_i=0$. As the bias voltage is increased, the cooling power increases as well. However, from a certain bias voltage on, the cooling power starts to decrease and finally turns negative for any value of level positions.}
\end{figure}

We first analyze the behavior of the heat flow driven at the third contact by the voltage difference $V=V_1-V_2$ applied between contacts 1 and 2
as a function of the dot level positions $E_L$ and $E_R$.
In what follows, we assume that each dot in Fig.~\ref{fig:nonlinear}(a)
couples symmetrically to the edge states on its left and right side.
This assumption does not affect the main conclusions of our work.
Figure~\ref{fig:NLHeatCurrent} displays the results in the isothermal case, i.e., when all terminals are kept at the same temperature ($\Delta T_i=0$).
For small voltages $eV_i\ll k_BT$, the system works as a 
refrigerator that cools terminal $3$ since $I_3^h>0$. This situation occurs whenever the dot level positions are tuned such that $E_R>0$ 
and $E_L<0$. In fact, the maximal cooling power occurs for $E_R=-E_L\simeq 2.5k_BT$ for the chosen parameters, see Fig.~\ref{fig:NLHeatCurrent}(a).
As the bias voltage is increased, the absolute value of the heat current 
grows as well as shown in Fig.~\ref{fig:NLHeatCurrent}(b). However, this increase is stronger for $E_L=-E_R>0$, in which case the third terminal is always heated ($I_3^h<0$),
than for the cooling condition $E_L=-E_R<0$.  By further increasing the bias voltage [Fig.~\ref{fig:NLHeatCurrent}(c)], the region where the  cooling 
phenomenon occurs  disappears and the third lead only gets heated for any configuration of the dot levels. This is originated from the fact that
for large values of $V$ electrons at 
terminal $1$ possess a high potential energy and are consequently transmitted to terminal $3$, leading to a heating of the third terminal.
For this to occur two aspects are crucial: on one hand, the fact that the dot resonances are voltage shifted. The level renormalization is due to Coulomb repulsion
and this further supports the importance of including interactions in the nonlinear regime of transport. On the other hand, we have considered finite resonance lifetimes $\Gamma\gtrsim k_{\rm B}T$ such that at large enough voltage, the injection (absorption) of electrons above (below) the chemical potential of terminal 3 are favorable, hence limiting the performance of a realistic refrigerator.

Remarkably, the chirality of the electron motion introduces an additional asymmetry in the device, as electrons injected from terminal 3 are absorbed by terminal 2 conditioned on being reflected first at quantum dot $L$, while electrons injected from 2 are never absorbed by 1. This allows our system to work as a refrigerator even for mirror symmetric configurations with $E_L=E_R$ and $\Gamma_L=\Gamma_R$, as can be clearly seen in the diagonal of Fig.~\ref{fig:NLHeatCurrent}(a). (Note that standard thermocouples rely on simultaneous broken electron-hole and mirror symmetries~\cite{benenti_fundamental_2017}.) Nonlinear voltages reduce and eventually removes this effect, see Fig.~\ref{fig:NLHeatCurrent}(b) and (c).

\begin{figure}
	\centering
	\includegraphics[width=.5\textwidth]{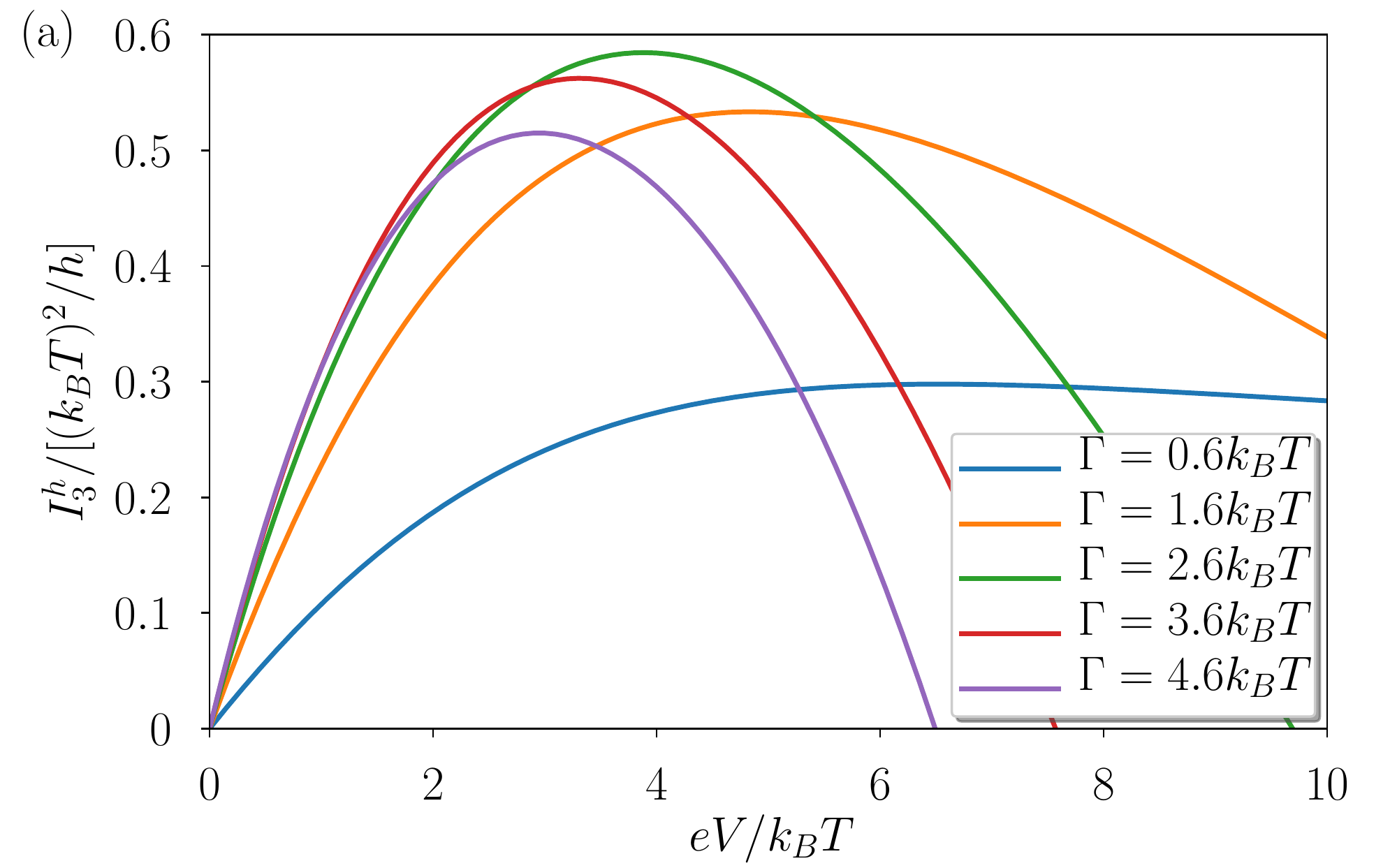}
    \includegraphics[width=.5\textwidth]{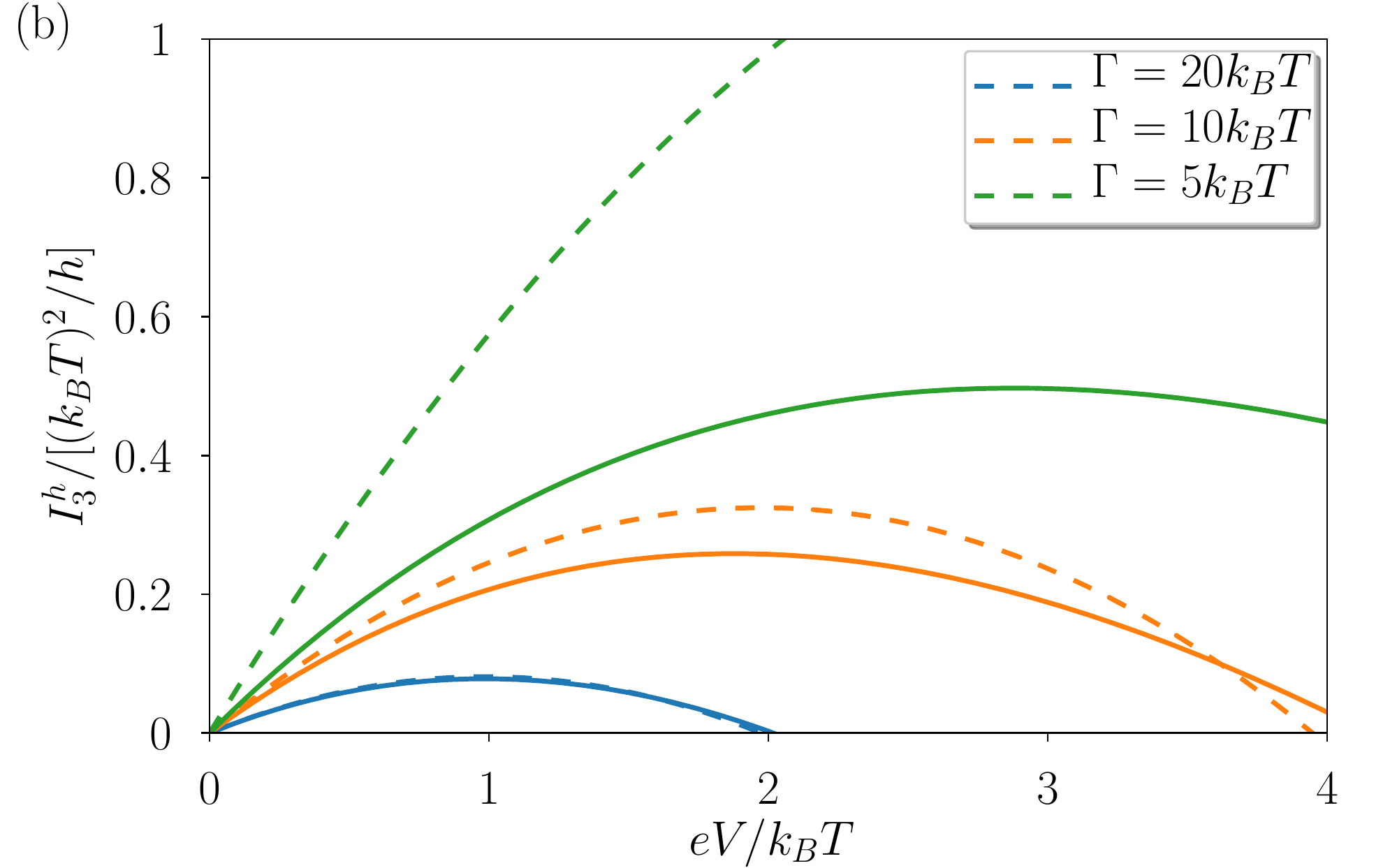}
	\caption{\label{fig:NLPerformance}(a) Cooling power at $\Delta T_i=0$ as a function of the applied bias voltage for different level widths. Level positions are $E_R=-E_L=2.9k_BT$. (b) Heat current flowing out of terminal 3 for different linewidths. Solid lines correspond to the numerical solution while dashed lines refer to the analytical solution. We set the level positions as $E_\text{R}=-E_\text{L}=\Gamma/2$.}
\end{figure}

Let us now take a closer look at the refrigerator operation in the nonlinear regime.
Figure~\ref{fig:NLPerformance}(a) shows the cooling power of the refrigerator $I_3^h$ as a function of the applied voltage for different values of the dot level broadening.  Quite generically, we find that the maximum cooling power is a nonmonotonic function of the voltage difference. We observe that the heat current in terminal 3 first rises as $V$ is increased, then reaches a maximum (the maximum cooling power) and finally drops for larger voltages. We note that the bias voltage at which the maximal cooling power occurs decreases as the level width $\Gamma=\Gamma_L=\Gamma_R$ enhances. Similarly, the dependence of the maximal cooling power on the level width is nonmonotonic. It attains its highest value when  $\Gamma\simeq 2.6k_BT$. This optimal value is related to the dots acting as energy filters but at the same time with sufficiently large values of energy broadening that allow for nonnegligible heat flows. 

We now show that our numerical findings can be understood within a weakly nonlinear transport model~\cite{lopez_nonlinear_2013}. This model is based on a perturbative expansion in terms of both electric and thermal gradients. We focus on an analysis of the cooling performance. Hence, it suffices for our purposes to expand the heat current  $I_3^h$ up to second order in the applied bias voltage. In this fashion, the cooling power reads
\begin{equation}\label{eq:I3h}
I_3^h = \sum_\alpha R_{3\alpha} V_\alpha + 
\sum_{\alpha\beta} R_{3\alpha\beta} V_\alpha V_\beta\,,
\end{equation}
where $\alpha$ and $\beta$ are lead indices,
$R_{3\alpha} = (\partial I_3^h /\partial V_\alpha)_{\rm eq}$ and
$R_{3\alpha\beta} = (1/2) (\partial^2 I_3^h /\partial V_\alpha\partial V_\beta)_{\rm eq}$
are the linear and leading-order nonlinear responses, respectively,
evaluated at equilibrium (i.e., for $V_1=V_2=V_3=0$) \cite{lopez_nonlinear_2013}.
Using the sum rules enforced by unitarity of the scattering matrix,
$\sum_{\beta} R_{\alpha\beta}
=\sum_{\beta\gamma} (R_{\alpha\beta\gamma}-R_{\alpha\gamma\beta})=0$ \cite{meair_scattering_2013}, Eq.~\eqref{eq:I3h} becomes
\begin{eqnarray}
I_3^h &=& R_{31}(V_1 - V_3) + R_{32}(V_2 - V_3) - R_{312}(V_1-V_2)^2\nonumber
\\
&& - R_{313}(V_1 - V_3)^2 - R_{323}(V_2 - V_3)^2 \,. \label{eq:I3h2}
\end{eqnarray}
We stress that this formalism leads to gauge-invariant expressions 
unlike alternative approaches~\cite{wang_nonlinear_2018} since 
electron-electron interactions are included. We consider that terminal 3 is a floating probe,
whose electrochemical potential is determined from the condition that there is no net flow of particles through this terminal:  $I_3^e=0$.
To be consistent,
this charge current is expressed up to second order in a bias voltage expansion,
\begin{equation}\label{eq:I3e}
I_3^e = \sum_\alpha G_{3\alpha} V_\alpha + 
\sum_{\alpha\beta} G_{3\alpha\beta} V_\alpha V_\beta\,,
\end{equation}
Here, $G_{3\alpha} = (\partial I_3^e /\partial V_\alpha)_{\rm eq}$ and
$G_{3\alpha\beta} = (1/2) (\partial^2 I_3^e /\partial V_\alpha\partial V_\beta)_{\rm eq}$ are the linear and weakly nonlinear electrical conductances. Substituting $V_3$ into Eq.~\eqref{eq:I3h2} we find an analytical expression for the cooling power as a function of the transport coefficients,
\begin{align}\label{eq_I3h2}
I_3^h &=\frac{G_{32}R_{31}-G_{31}R_{32}}{G_{31}+G_{32}}V
-\frac{R_{313}G_{32}^2+R_{323}G_{31}^2}{(G_{31}+G_{32})^2}V^2 \nonumber\\
&-R_{312}V^2+\frac{R_{31}+R_{32}}{(G_{31}+G_{32})^3}
\left[G_{312}(G_{31}+G_{32})^2\right.\nonumber\\
&\left.+G_{313}G_{32}^2+G_{323} G_{31}^2\right]V^2\,,
\end{align}
where we have set $V_1=-V_2=V/2$ without loss of generality because, as discussed before,
our theory is gauge invariant~\cite{sanchez_scattering_2013}. A simpler expression can be
obtained for $I_3^h$ when the following assumptions are made. First,
at very low temperature the coefficients $R_{\alpha\beta\gamma}$  are proportional
to $\delta_{\alpha\beta}+ \delta_{\alpha\gamma}-\delta_{\beta\gamma}$~\cite{lopez_nonlinear_2013}. Therefore,
we can neglect the third term in the right-hand side of Eq.~\eqref{eq_I3h2}. Second,
the coefficients $R_{\alpha\beta}$ and $G_{\alpha\beta\gamma}$ depend
on the energy derivative of the transmission~\cite{lopez_nonlinear_2013}. As a consequence, the last term
in the right-hand side of Eq.~\eqref{eq_I3h2} is a small correction in the low temperature limit
and Eq.~\eqref{eq:I3h2} simplifies to
\begin{align}\label{eq_I3h2b}
I_3^h =\frac{G_{32}R_{31}-G_{31}R_{32}}{G_{31}+G_{32}}V
-\frac{R_{313}G_{32}^2+R_{323}G_{31}^2}{(G_{31}+G_{32})^2}V^2\,,
\end{align}
Interestingly, our problem effectively becomes noninteracting since the
nonlinear responses in Eq.~\eqref{eq_I3h2b} are quite insensitive to variations of the scattering
matrix with voltage and therefore to nonlinear screening. This can be seen if we apply a Sommerfeld expansion:
\begin{align} \label{eq_R31}
R_{31}&=-\frac{2\pi^2 ek_B^2 T^2}{3h} \mathcal T_{31}'\,,\\
\label{eq_R32}R_{32}&=-\frac{2\pi^2 ek_B^2 T^2}{3h} \mathcal T_{32}'\,,\\
\label{eq_R313}R_{313}&=\frac{e^2}{h}\mathcal T_{31}=-\frac{1}{2}G_{31}\,,\\
\label{eq_R323} R_{323}&=\frac{e^2}{h}\mathcal T_{32}=-\frac{1}{2}G_{32},,
\end{align}
where the energy is evaluated at $E_F$ and primes indicate energy derivatives.

Equation~\eqref{eq_I3h2b} is able to qualitatively explain the behavior
of the cooling power shown in Fig.~\ref{fig:NLPerformance}(a).
Cooling operation mode switches on when $G_{32}R_{31}<G_{31}R_{32}$
since the second term in the right-hand side of Eq.~\eqref{eq_I3h2b}
is always negative. 
For our refrigerator setup in Fig.~1, we have
$\mathcal T_{31}=\mathcal{T}_L(1-\mathcal{T}_R)$ and $\mathcal T_{32}=\mathcal{T}_R$, where $\mathcal{T}_L$ ($\mathcal{T}_R$)
is the transmission probability through the left (right) dot in Fig.~\ref{fig:nonlinear}.
Therefore, cooling demands that 
\begin{equation}
\label{eq:transmcool}
\frac{\mathcal{T}_L'}{\mathcal{T}_L} < \frac{\mathcal{T}_R'}{\mathcal{T}_R(1-\mathcal{T}_R)}\,,
\end{equation}
where the functions $\mathcal{T}_{L/R}$ and their energy derivatives $\mathcal{T}'_{L/R}$ are evaluated at the Fermi energy.
This condition is translated into the fact that $E_L<E_R$ is most favorable to have cooling. This precisely is what has been numerically found
in Fig.~\ref{fig:NLHeatCurrent}. However, the asymmetry between left and right hand sides of Eq.~\eqref{eq:transmcool}, which is once more a consequence of chirality, allows for cooling even if ${\cal T}_L={\cal T}_R$.

A more physical and intuitive interpretation can be given if we recast $G_{32}R_{31}<G_{31}R_{32}$
as $S_{31}>S_{32}$, where $S_{ij}\propto d\ln T_{ij}/dE$ for $E=E_F$
is a thermopower or Seebeck coefficent generalized to multiterminal
setups~\cite{matthews_thermally_2012,matthews_experimental_2014,michalek_local_2016}.
Then, for $E_L=-E_R<0$ the right dot supports electrons traveling above the Fermi energy (i.e., the thermopower is negative)
while the left dot supports electrons traveling below the Fermi energy (i.e., the thermopower is positive),
which is consistent with the condition $S_{31}>S_{32}$.
The quasiparticle energy difference is supplied by the terminal which is being cooled. 
 
Furthermore, Eq.~\eqref{eq_I3h2b} predicts that the maximum cooling power $I_{3,m}^h$
will occur at the applied voltage
\begin{align}
V_m = \frac{1}{2}\frac{(G_{31}+G_{32})(G_{32}R_{31}-G_{31}R_{32})}
{R_{313}G_{32}^2+R_{323}G_{31}^2}\,,
\end{align}
in which case 
\begin{align}\label{eq:I3mh}
I_{3,m}^h = \frac{1}{4}\frac{(G_{32}R_{31}-G_{31}R_{32})^2}
{R_{313}G_{32}^2+R_{323}G_{31}^2}\,.
\end{align}
Since this expression involves energy derivatives of the dot transmissions,
it is natural to expect that $I_{3,m}^h$ becomes optimal when the transmissions
change more rapidly, i.e., for energies of the order of $\Gamma$
(maximal charge fluctuations), in agreement
with the full numerical calculations. Therefore, quantum dot junctions
with large thermopowers will show higher cooling powers but with proper
tuning of the tunnel broadening value as we already showed in our numerics. Indeed, for very small $\Gamma$, 
$R_{31}$ (as the transmission derivative) increases but $G_{32}$ decreases.
Therefore, there exists an optimal value of $\Gamma$ that maximizes 
$I_{3,m}^h$, explaining the nonmonotonic behavior of $I_{3,m}^h$ versus $\Gamma$
depicted in Fig.~\ref{fig:NLPerformance}(a). A comparison between the weakly nonlinear analysis
and the exact solution is shown in Fig.~\ref{fig:NLPerformance}(b). The agreement in the low temperature
case is excellent ($\Gamma=20 k_B T$). Deviations are stronger for smaller ratios of $\Gamma/k_B T$ since
the Sommerfeld expansion of Eqs.~\eqref{eq_R31}, \eqref{eq_R32}, \eqref{eq_R313}, and~\eqref{eq_R323} breaks down.
However, the weak coupling limit ($\Gamma\ll k_B T$) is less interesting from a practical point of view because
it leads to lower cooling powers [see, e.g., the blue curve in Fig.~\ref{fig:NLPerformance}(a)].

\begin{figure}
	\includegraphics[width=.5\textwidth]{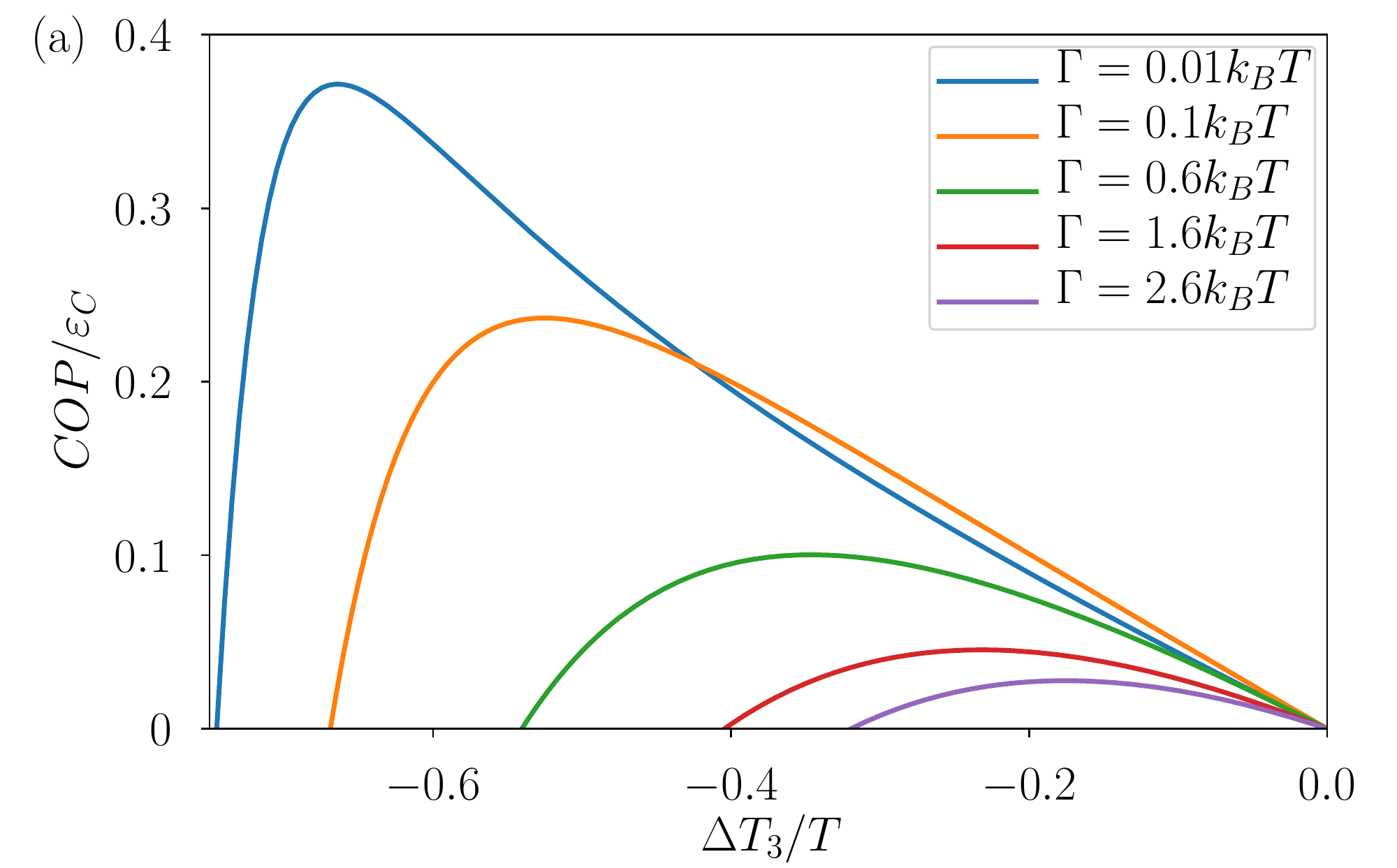}
	\includegraphics[width=.5\textwidth]{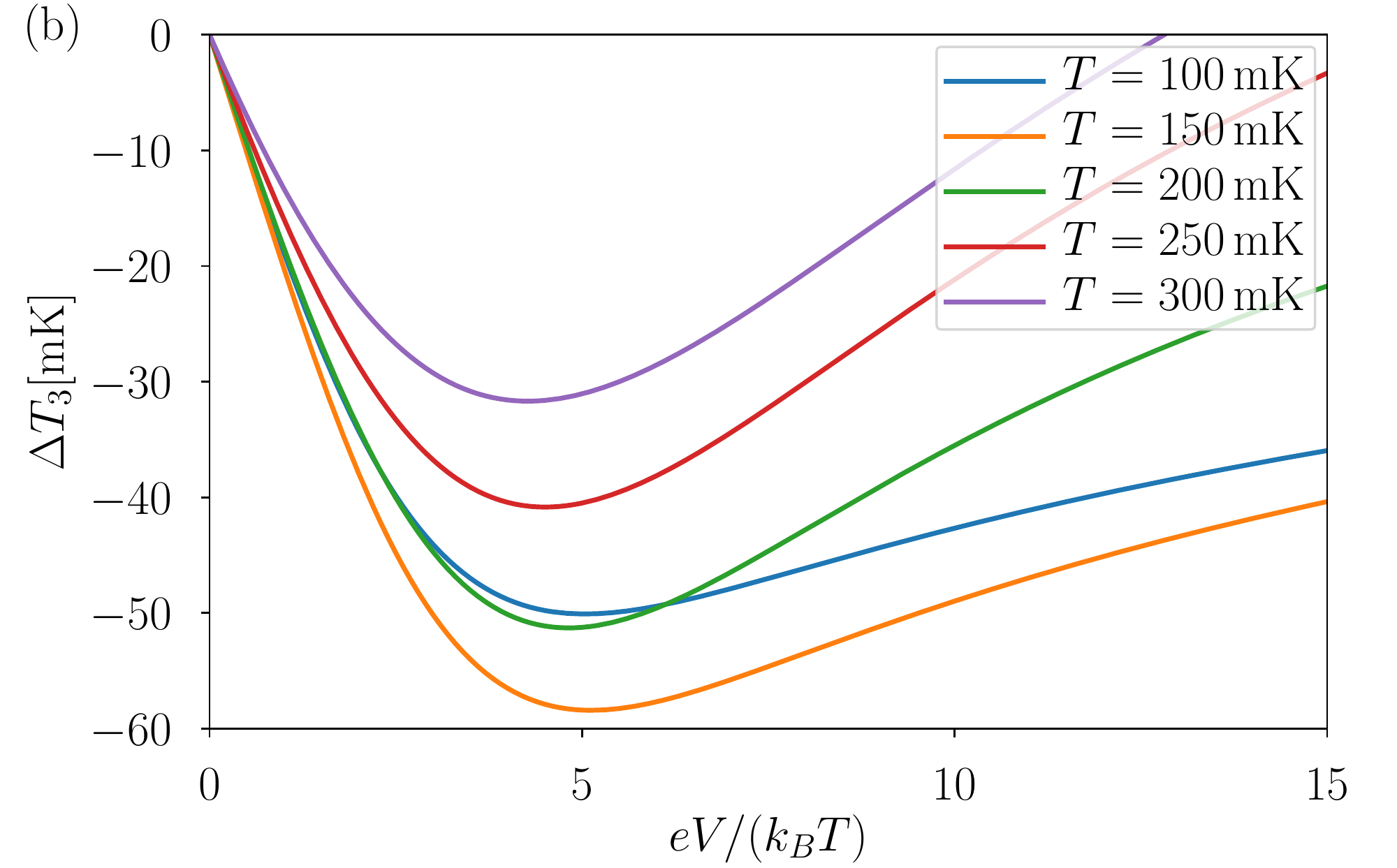}
	\caption{\label{fig:comparison} 
	(a) Coefficient of performance in units of the Carnot efficiency $\varepsilon_C=-(T+\Delta T_3)/\Delta T_3$ as a function of the temperature bias for different level widths. For each curve the bias voltage has been chosen such as to generate the largest possible heat current at zero temperature difference. Level positions are chosen as $E_R=-E_L=2.9k_BT$. (b) Largest temperature difference that the refrigerator can reach in the presence of a phonon bath for different base temperatures with a symmetrically applied bias voltage $V=V_1/2=-V_2/2$. For each base temperature level width and level position are optimized to create the largest possible temperature difference. Parameters are $\Sigma=\unit[10^9]{WK^{-5}m^{-3}}$ and $\mathcal V=\unit[10^{-20}]{m^3}$. }
\end{figure}

Next, we discuss the efficiency of our refrigeration process. This is quantified by the
coefficient of performance  
\begin{equation}
{\rm COP}=\frac{I^h_3}{I^e_1(V_1-V_2)}\,.
\end{equation}
In Fig.~\ref{fig:comparison}(a) we show COP expressed in units of the Carnot efficiency $\varepsilon_C=-(T+\Delta T_3)/\Delta T_3$ as a function of the temperature at the third lead, $\Delta T_3/T$, for various values of the dot level broadening. The optimal configuration for the cooling performance is obtained when the level broadening $\Gamma$ is small and the temperature difference is sufficiently large. This case in Fig.~\ref{fig:comparison}(a) is represented by the curve corresponding to $\Gamma=0.01 k_BT$ when $\Delta T_3/T\simeq 0.5$. As a general trend, we find that as the temperature gradient is increased, the COP grows, reaches a maximum and finally drops to zero for very large $\Delta T_3$. For broad resonances, the COP shows a nearly quadratic dependence on $\Delta T_3$. As the level width is decreased, this quadratic dependence no longer holds and the maximal COP takes place for smaller values of $T_3$. We furthermore remark that decreasing the level width strongly enhances the COP. This behavior can be understood as follows. For sharp levels, the system has excellent energy filtering properties that lead to very efficient cooling. At the same time, the heat and charge currents that can flow in the device are small due to the smallness of $\Gamma$ that diminishes the input power $I^e_1(V_1-V_2)$. However, for broad levels energy filtering becomes poor, thus degrading the performance of the refrigerator although the input power increases. In between these two configurations, there is an optimal level width that gives rise to a large cooling power with a relatively small input power in combination with a reasonable COP. This is similar to what was found previously for heat engines based on resonant tunneling through quantum dots~\cite{jordan_powerful_2013} and quantum wells~\cite{sothmann_powerful_2013}. 

A further important quantity to characterize the refrigeration performance is the maximal temperature difference that the cooler can generate (i.e., the value of $\Delta T_3$ at which $I^h_3=0$ for a given bias voltage). In order to determine this quantity, we consider a simple thermal model where terminal 3 couples to phonons in the substrate giving rise to a heat current due to electron-phonon coupling
\begin{equation}
	P_\text{e-ph}=\Sigma\mathcal{V} (T_3^5-T_\text{ph}^5)
\end{equation}
which again is positive if heat flows from the electric conductor into the phonon bath. In the above expression, $\Sigma$ characterizes the strength of the electron-phonon coupling. For typical metals, it takes the value $\Sigma\simeq \unit[10^9]{WK^{-5}m^{-3}}$. $\mathcal V$ is the volume of the conductor which for a typical mesoscopic structure is of the order of $\mathcal{V}\simeq \unit[10^{-20}]{m^3}$. Finally, $T_\text{ph}$ denotes the temperature of the phonon bath.
The stationary temperature of terminal 3 is determined by the balance equation
\begin{equation}\label{eq:balance}
	I^h_3+P_\text{e-ph}=0.
\end{equation}
As $I^h_3$ scales quadratically with the temperature while the phononic contribution scales as $T^5$, we immediately conclude that the  refrigerator works best at low temperatures. This can be seen in Fig.~\ref{fig:comparison}(b), which shows the minimal temperature that the refrigerator can reach at different base temperatures. As temperature is lowered, the relative change in temperature increases. In terms of the absolute temperature difference that can be obtained, we find that it occurs in the temperature range between $100$ and \unit[200]{mK}. 

We can also apply our previous weakly nonlinear model to discuss the minimum possible temperature.
Substituting Eq.~\eqref{eq_I3h2b} into Eq.~\eqref{eq:balance}, we can
solve for the minimum attainable temperature,
\begin{align}
(\Delta T_3)_{\rm min} = T \left(
1 + \frac{a^2}{4b\Sigma\mathcal{V} T^5}
\right)^{1/5}\,,
\end{align}
where $a=(G_{32}R_{31}-G_{31}R_{32})/(G_{31}+G_{32})$
and $b=(R_{313}G_{32}^2+R_{323}G_{31}^2)/(G_{31}+G_{32})^2$.
Importantly, the minimum temperature increases with $a$,
which is governed by the thermopower asymmetry between
the dots, $S_{31}-S_{32}$, much like the maximum cooling power given
by Eq.~\eqref{eq:I3mh}. This explains why the optimum voltage value
is similar in both Figs.~\ref{fig:NLPerformance}(a) and~\ref{fig:comparison}(b).

We would like to point out that in the absence of phonons the refrigerator would achieve lowest temperatures with very narrow resonances because in this case cooling is highly efficient. As the absolute cooling power is very small in this case, even a very weak electron-phonon coupling gives rise to heat currents from the substrate into the electrode which are much larger than the cooling power. Hence, in order to operate the device efficiently in a realistic scenario, level widths of the order of $k_BT$ are required.

\section{\label{sec:conclusions}Conclusions}
In closing, we propose a refrigerator based on chiral quantum Hall edge channels present in a three-terminal conductor hosting two quantum dot scatterers. The device operates by driving a charge current between terminals 1 and 2. As a consequence, heat is extracted from terminal 3 that acts as a voltage probe.
Our transport theory fully accounts for nonlinear effects. Electron-electron interactions are included at the mean-field level, which is a good approximation in the absence of strong correlations. We have found that cooling is possible only by means of certain energy-dependent scattering  that in our setup is provided by the two quantum dots.  The advantage of this multiterminal configuration is that there is no direct Joule heating in the cooled terminal.  Our findings indicate that by properly tunning the dot level couplings to the edge states the setup can work as a powerful and efficient refrigerator with distinctive features arising from chirality. We estimate that for realistic devices our refrigerator can cool by about \unit[60]{mK} at a base temperature of \unit[150]{mK}.
For comparison, Ref.~\cite{feshchenko_experimental_2014} reaches with a Coulomb blockade refrigerator
a maximum value of \unit[15]{mK} at a base temperature of \unit[90]{mK}.
Therefore, our results might be of great importance for the implementation of coolers in nanochips
simultaneously showing optimal performance and avoiding heating effects.

\begin{acknowledgments}
We acknowledge financial support from the Ministry of Innovation NRW via the ``Programm zur Förderung der Rückkehr des hochqualifizierten Forschungsnachwuchses aus dem Ausland'', the MICINN Grant No.\ MAT2017-82639, the Spanish MINECO via grant No. FIS2015-74472-JIN (AEI/FEDER/UE), the Ram\'on y Cajal program RYC-2016-20778, the ``Mar\'ia de Maeztu\textquotedblright{} Program for Units of Excellence in R\&D (MDM-2014-0377 and MDM-2017-0711), and the MAT2016-82015-REDT network.
\end{acknowledgments}

\appendix

\section{Discretization}\label{sec:discretization}
The different regions of constant potential, $A_1$, $A_2$, $B_1$, $B_2$, $B_3$, $C_1$, $C_2$, $\Omega_L$ and $\Omega_R$, are shown in Fig.~\ref{fig:nonlinear}(a). In the following, we assume that all edge channels have the same, constant density of states $\nu_\text{F}$. The transmission and reflection amplitudes of the two quantum dots are given by
\begin{align}
	t_r&=\frac{1}{2}\left(e^{2i\delta_r}-1\right)\,,\\
	r_r&=\frac{1}{2}\left(e^{2i\delta_r}+1\right)\,,
\end{align}
with
\begin{equation}
	e^{2i\delta_r}=\frac{E-E_r-i\Gamma_r/2}{E-E_r+i\Gamma_r/2}\,.
\end{equation}
The corresponding transmission probability then reads 
\begin{equation}
	\mathcal{T}_r=1-\mathcal{R}_r=|t_r|^2=\frac{\Gamma_r^2/4}{(E-E_r-eU_{\Omega r})^2+\Gamma_r^2/4}\,.
\end{equation}
The particle injectivities $\nu^p_{X,j}$ of lead $j$ in region $X$ are then given by
\begin{align*}
	\bar \nu^p_{A1,1}&=\nu_\text{F},& \bar \nu^p_{A1,2}&=0,& \bar \nu^p_{A1,3}&=0,\\
	\bar \nu^p_{A2,1}&=\mathcal{R}_L \nu_\text{F},& \bar \nu^p_{A2,2}&=0,& \bar \nu^p_{A2,3}&=\mathcal{T}_L \nu_\text{F}, \\
	\bar \nu^p_{B1,1}&=0,& \bar \nu^p_{B1,2}&=0,& \bar \nu^p_{B1,3}&=\nu_\text{F}, \\
	\bar \nu^p_{B2,1}&=\mathcal{T}_L \nu_\text{F},& \bar \nu^p_{B2,2}&=0,& \bar \nu^p_{B2,3}&=\mathcal{R}_L \nu_\text{F}, \\
	\bar \nu^p_{B3,1}&=\mathcal{T}_L \mathcal{R}_R \nu_\text{F},& \bar \nu^p_{B3,2}&=\mathcal{T}_R \nu_\text{F},& \bar \nu^p_{B3,3}&=\mathcal{R_L} \mathcal{R_R} \nu_\text{F}, \\
	\bar \nu^p_{C1,1}&=0,& \bar \nu^p_{C1,2}&=\nu_\text{F},& \bar \nu^p_{C1,3}&=0, \\
	\bar \nu^p_{C2,1}&=\mathcal{T}_L \mathcal{T}_R \nu_\text{F},& \bar \nu^p_{C2,2}&=\mathcal{R}_R \nu_\text{F},& \bar \nu^p_{C2,3}&=\mathcal{R_L} \mathcal{T_R} \nu_\text{F}.
\end{align*}
The injectivities for the two quantum dots are given by
\begin{align}
	\nu^p_{\Omega L,1}&=-\frac{1}{2\pi i}\left(r_L^* \frac{\delta r_L}{e\delta U_{\Omega L}}+t_L^* \frac{\delta t_L}{e\delta U_{\Omega L}}\right)\\
	&=+\frac{1}{2\pi i}\left(r_L^* \frac{d r_L}{dE}+t_L^* \frac{d t_L}{dE}\right)\\
	&=\frac{1}{\pi}\frac{\Gamma_L/4}{(E-E_L-eU_{\Omega L})^2+\Gamma_L^2/4}=\frac{\mathcal{T}_L}{\pi \Gamma_L}\,,
\end{align}
and similarly
\begin{equation}
	\nu^p_{\Omega L,2}=0\,,
\end{equation}
and
\begin{equation}
	\nu^p_{\Omega L,3}=\frac{1}{\pi}\frac{\Gamma_L/4}{(E-E_L-eU_{\Omega L})^2+\Gamma_L^2/4}=\frac{\mathcal{T}_L}{\pi \Gamma_L}\,,
\end{equation}
as well as
\begin{equation}
	\nu^p_{\Omega R,2}=\frac{1}{\pi}\frac{\Gamma_R/4}{(E-E_R-eU_{\Omega R})^2+\Gamma_R^2/4}=\frac{\mathcal{T}_R}{\pi \Gamma_R}\,.
\end{equation}
For the injectivities $\nu^p_{\Omega R,1}$ and $\nu^p_{\Omega R,3}$ we need to take into account the presence of the left dot which gives rise to an additional factor $\mathcal{T}_L$ and $\mathcal{R}_L$, respectively:
\begin{align}
	\nu^p_{\Omega R,1}&=\frac{\mathcal{T}_L}{\pi}\frac{\Gamma_R/4}{(E-E_R-eU_{\Omega R})^2+\Gamma_R^2/4}=\frac{\mathcal{T}_L \mathcal{T}_R}{\pi \Gamma_R}\,,\\
	\nu^p_{\Omega R,3}&=\frac{\mathcal{R}_L}{\pi}\frac{\Gamma_R/4}{(E-E_R-eU_{\Omega R})^2+\Gamma_R^2/4}=\frac{\mathcal{R}_L \mathcal{T}_R}{\pi \Gamma_R}\,.
\end{align}

\section{Integrals}\label{sec:integrals}

In order to evaluate the nonequilibrium charge, we need the following integrals:
\begin{align}
	\mathcal{I}_1^r(\mu_j,T_j)&=\int \frac{dE}{(E-E_r)^2+\Gamma_r^2/4}f_j(E)\nonumber\\
	&=\frac{\pi}{\Gamma_r}-\frac{2}{\Gamma_r}\im \Psi(z_j^+)\,,
\end{align}
where $\Psi(z_j^+)$ is the digamma function with argument $z_j^\pm=\frac{1}{2}+\frac{\Gamma_r}{4\pi k_B T_j}\pm\frac{E_r-\mu_j}{2\pi k_B T_j}$, and
\begin{align}
\mathcal{I}_2(\mu_j,T_j) &=
\int \frac{f_j(E) dE}{[(E-E_L)^2+\Gamma_L^2/4][(E-E_R)^2+\Gamma_R^2/4]}\nonumber\\
&=\frac{\Gamma_L{+}\Gamma_R}{\Gamma_L\Gamma_R}\frac{\pi}{\eta}
	-\frac{2}{\eta^2-\Gamma_L^2\Gamma_R^2/4} \nonumber\\
&\times\Bigl\{(E_L{-}E_R)\left[\re\Psi(z_{Lj}^+){-}\re\Psi(z_{Rj}^+)\right]\nonumber\\
	&+\sum_{r\neq r'} \frac{\Gamma_r}{\Gamma_1\Gamma_2}\left[(E_L{-}E_R)^2{+}\frac{\Gamma_{r'}^2{-}\Gamma_r^2}{4}\right]\im\Psi(z_{rj}^+)\Bigr\}\,,
\end{align}
where $z_{rj}^\pm=\frac{1}{2}+\frac{\Gamma_r}{4\pi k_BT_j}\pm\frac{E_r-\mu_j}{2\pi k_BT_j}$ and $\eta=(E_L-E_R)^2+(\Gamma_1^2+\Gamma_2^2)/4$.

\bibliographystyle{apsrev4-1}
\bibliography{Meine_Bibliothek}
 

\end{document}